\documentclass[10pt]{article}
\usepackage{amssymb}
\usepackage{epsfig}
\input epsf
\begin{document}
\hyphenation{ge-ne-ra-tes}
\hyphenation{me-di-um  as-su-ming pri-mi-ti-ve pe-ri-o-di-ci-ty}
\hyphenation{mul-ti-p-le-sca-t-te-ri-ng i-te-ra-ti-ng e-q-ua-ti-on}
\hyphenation{wa-ves di-men-si-o-nal ge-ne-ral the-o-ry sca-t-te-ri-ng}
\hyphenation{di-f-fe-r-ent tra-je-c-to-ries e-le-c-tro-ma-g-ne-tic pho-to-nic}
\title{
\hfill{}\\
\hfill{}\vspace*{0.5cm}\\
\sc
Photonic band gaps of three-dimensional face-centered cubic lattices
\vspace*{0.3cm}}
\author{ {\sc Alexander Moroz}${}^1$\thanks{e-mail address :
moroz@amolf.nl}
%
\, and {\sc Charles Sommers}${}^{2}$\thanks{e-mail address :
root@classic.lps.u-psud.fr}
\vspace*{0.3cm}}
\date{
\protect\normalsize
\it ${}^1$ FOM-Instituut voor  Atoom- en Molecuulfysica, Kruislaan 407, 
1098 SJ Amsterdam, 
The Netherlands\\
\it ${}^2$ Laboratoire de Physique des Solides,
Univ. Paris-Sud, B\^{a}timent 510, F-91405 Orsay Cedex, France}
\maketitle
\begin{center}
{\large\sc abstract}
\end{center}
We  show that the photonic analogue of the Korringa-Kohn-Rostocker 
method is a viable alternative to the plane-wave method 
to analyze the spectrum of electromagnetic waves in a three-dimensional 
periodic dielectric lattice. Firstly, in the case of an fcc lattice 
of homogeneous dielectric spheres,
we reproduce the main features of the spectrum obtained by the plane 
wave method, namely that for a sufficiently high dielectric contrast
a full gap opens in the spectrum between the eights and ninth bands
if the dielectric constant $\varepsilon_s$ of spheres
is lower than the dielectric constant $\varepsilon_b$ of the background medium.
If $\varepsilon_s> \varepsilon_b$, no  gap is found in the spectrum.
The maximal value of the relative band-gap width approaches $14\%$ in 
the close-packed case and  decreases monotonically 
as the filling fraction decreases. The lowest dielectric contrast 
$\varepsilon_b/\varepsilon_s$ for which a full gap opens in the spectrum  
is  determined to be $8.13$. Eventually, in the case of an fcc lattice 
of coated spheres, we demonstrate that a suitable coating can enhance 
gap widths by as much as $50\%$.

\vspace*{0.6cm}

\noindent PACS numbers:  42.70.Qs, 71.20. \hfill


\vspace*{1.9cm}

\begin{center}
({\bf to appear in J. Phys.: Condens. Matter})
\end{center}

\thispagestyle{empty}
\baselineskip 20pt
\newpage
\setcounter{page}{1}
\section{Introduction}
Under certain conditions, a gap can open in the spectrum of 
electromagnetic waves in a dielectric medium, independent of the
direction of their propagation \cite{Y,Jo}. 
Dielectric structures possessing such a photonic band gap are promising 
candidates for various technological applications \cite{Y}.
Moreover, such structures offer a new laboratory to study various 
atomic processes. Indeed, if a gap opens in the spectrum
of electromagnetic waves, all parameters and characteristics of atom
placed in such a medium,
such as, for example,  atomic radius and its spontaneous emission rates
are expected to change.

In order to open such a gap, one considers
Maxwell's equations in a dielectric with a spatially periodic dielectric
function, in full analogy to the Schr\"{o}dinger equation
with a periodic potential \cite{Y,Jo}. In the latter case, the spectrum can be 
classified according to the Bloch momentum ${\bf k}$. Energy (frequency) 
levels $\nu_n$ are continuous functions of the Bloch momentum ${\bf k}$
in the (first) Brillouin zone. 
We say that there is a full gap, or simply a {\em gap}
 between the $n$th and $(n+1)$th 
levels when $\nu_{n+1}({\bf k})>\nu_n({\bf k}')$ for all  ${\bf k}$ 
and ${\bf k}'$. We say that there is a {\em direct gap} between the $n$th 
and $(n+1)$th levels when $\nu_{n+1}({\bf k})>\nu_n({\bf k})$ for all  
${\bf k}$.
For the Schr\"{o}dinger equation in one space dimension, the number 
of gaps is in general infinite and the only periodic potential which
does not open any gap in the spectrum is a constant potential
\cite{E}. However, the situation changes dramatically in two and 
higher dimensions. One can prove rigorously
that the number of gaps  in the spectrum can only  be  finite   
and, if the  potential is not strong enough, no gap opens in the 
spectrum \cite{S}. If electromagnetic waves are considered,
opening a  gap in the spectrum is even 
more difficult and it took several years of intensive
search to achieve it experimentally for microwaves \cite{YGL}.
Note that Maxwell's equations enjoy scale invariance so that, in principle,
by scaling all sizes of a given structure one can shift 
a gap theoretically to whatever frequency range. 

So far, the plane-wave method \cite{ZS,HCS,SHI,SY,BSS} has been the main tool 
to calculate  the spectrum of electromagnetic waves in three-dimensional 
dielectric lattices. However, the plane-wave method is numerically
rather unstable for a setup considered
in experiment, namely when the dielectric function is piece-wise
constant and changes discontinuously \cite{YGL,As,WV}.
The main culprit for this behaviour is the
Gibbs instability - the dielectric constant is poorly
estimated near its spatial discontinuities by a truncated Fourier series
which wildly oscillates
even if more than one thousand  of plane waves is retained
(see, for example, figure 2 of \cite{SHI}).
Also, the plane wave expansion becomes impractical if the dielectric
constant exhibits a large imaginary part. 
Another approach to calculate the spectrum 
of electromagnetic waves in three-dimensional 
dielectric lattices uses a discretization of Maxwell's equation
inside the primitive cell of the lattice \cite{Pen}.
However, both methods are difficult to apply in the presence  of impurities
and  to the calculation of  Green's function. 

In order to have an universal  method which can deal with
problems of the behaviour of electromagnetic waves in a periodic  
dielectric medium in their full complexity, we have developed and employed 
a photonic analogue of the first principle on-shell multiple-scattering  
theory (MST) and of the Korringa-Kohn-Rostocker (KKR) method \cite{Mo}. 
The unique feature of the on-shell MST is that, 
for nonoverlapping (muffin-tin) scatterers \cite{Bb}  
(the present situation), it disentangles single-scattering and 
multiple-scattering effects  (see \cite{MT} for a recent discussion).
The KKR method \cite{KKR} uses explicitly scattering matrices 
and Green's function  which are expanded in the basis 
of spherical harmonics and the spectrum is determined
by zeroes of a determinant. For electrons on a Bravais lattice, 
inclusion of spherical waves with angular momentum up to $l_{max}=2$ 
already
gives result within a few per cent of the exact calculation \cite{KKR}. 
Expansion in the basis of spherical harmonics does not mean that
scatterers have to be spherically symmetric. Indeed, scatterers
of arbitrary shape are allowed in which case scattering matrices
are simply nondiagonal in the angular momentum indices \cite{F1}.
The main advantage of the KKR method is that it gives directly
the total Green's function from which the density of states (DOS) and 
the so-called local density of states can be easily extracted. 
The local DOS, which is proportional to the imaginary part of the 
total Green's function at the coinciding points in the coordinate
space, is an important quantity  which determines decay of excited states 
of atoms and molecules embedded in the lattice \cite{STL}.
Also, the frequency dependence of the dielectric constant can be easily
implemented in the formalism.

The outline of our paper is as follows. In the following section,
we show that the photonic KKR method is a viable alternative to calculate 
the photonic band structure by  reproducing the main features of 
the spectrum  obtained by the plane-wave method. The lowest dielectric 
contrast 
$\varepsilon_b/\varepsilon_s$ for which a full gap opens in the spectrum  
is  found to be $8.13$ and is slightly lower than $8.4$
obtained by the plane-wave method \cite{BSS}.
In Section 3 we discuss the case of coated spheres, i.e.
spheres made out of several spherical shells with different
dielectric constant. We demonstrate that already a suitable 
single coating can enhance some of the gap widths by as much as $50\%$.
Our conclusions are summarized in Section 4.

\section{Face-centered cubic lattice of dielectric spheres}
In this section, we shall present the results of our numerical
calculation for a face-centered cubic (fcc) lattice of 
dielectric spheres with a single sphere per lattice
primitive cell.
This case is very interesting from the experimental 
point of view, since such dielectric lattices form
when silicon matrices, synthetic opals, and collodial crystals 
are used \cite{As,WV,TW}. Some of the structures were shown to 
exhibit the so-called {\em stop gap} (gap in the spectrum
at a fixed direction of the incident light)
at optical frequencies \cite{As,WV} and are the natural candidates
to achieve a full photonic band gap \cite{WV}.

At the same time, the case of
fcc lattice of dielectric spheres has been controversial since the first 
experimental results were published \cite{YG}. 
Results for a sample consisting from polycrystalline
$Al_2 O_3$ spheres, $6mm$ in diameter with a microwave refractive
index of $3.06$ in thermal-compression-molded dielectric
foam of refractive index $1.01$ indicated the presence
of a ``photonic band gap'' in the microwave regime \cite{YG}.
However, subsequent numerical 
calculations using the plane wave method \cite{ZS,HCS} claimed that no gap 
opens in the spectrum and only a pseudo-gap 
(a sharp drop in the DOS) exists \cite{ZS}.
Nevertheless, two years later  using the plane-wave method, 
 S\"{o}z\"{u}er,  Haus, and Inguva \cite{SHI} 
did find a full gap for the fcc lattice of dielectric spheres 
between the eights and ninth bands.
The discrepancy between the results of \cite{ZS,HCS} and \cite{SHI}
follows from the fact that unlike to the case of electrons,
a gap for electromagnetic waves opens in an intermediate region 
and the authors of \cite{ZS,HCS} stopped their
calculation just beneath that region (see figure 1 in \cite{ZS,HCS}).
Later, the results of S\"{o}z\"{u}er,  Haus, and Inguva \cite{SHI} 
were confirmed by two other groups using the plane-wave method
\cite{SY,BSS}.

The latter deserves some discussion.
In the case of electrons, the formation of bands results from the broadening 
of individual atomic levels when the atoms start to feel the presence 
of each other.  The largest gap between atomic levels is
between the lowest-lying energy levels. Therefore, for a lattice
of atoms, one expects to find a gap essentially between the first
and the second energy band with the gap  between
higher bands scaling down to zero \cite{E}. 
However,  for electromagnetic waves a gap does not open between the
lowest lying bands but in an intermediate region.
This phenomenon can be rather easily understood, since for a dielectric 
scatterer and Maxwell's equations bound states are absent. They are replaced 
by resonances  and the above argument for locating the position of a 
band gap no longer holds. Moreover, if the wavelength is small 
compared to the size of the spheres, one can use geometric optics while in the
opposite limit of long wavelengths, the Rayleigh approximation applies.
In neither case a gap opens in the spectrum.
Therefore, if a gap is present in the spectrum, it should be
in the intermediate region between the two limiting cases
(see, however, the case of a diamond lattice (\cite{HCS}, figure 2), 
which is a complex lattice).
The very same is also expected to apply for the localization
of light \cite{Jo,WL}.

\subsection{Results}
S\"{o}z\"{u}er,  Haus, and  Inguva \cite{SHI} were well aware of the 
convergence problems of the plane-wave method and they called
for the recalculation and confirmation of their results by a 
more precise method. The latter constitutes the first
part of our results. Using the photonic KKR method,
we were able to confirm the plane-wave method result \cite{SHI} 
that, in the case of air spheres and for a sufficiently 
high dielectric contrast
\begin{itemize}
\item  a full gap opens  between
the eights and ninth bands   
\item a direct gap opens between the fifth and sixth bands.
\end{itemize}
If the dielectric constant of spheres is larger than that of a 
background medium, no gap opens in the spectrum. This situation
is realized, for example, if dielectric spheres in air are considered.
We did not find any compelling explanation for this behaviour. 
In general, the higher frequency the higher $l_{max}$ is to
be taken. Taking $l_{max}=1$ is sufficient to account for the linear 
part of the band structure around the $\Gamma$ point.
The intermediate region requires then $l_{max}=3-5$ and 
$l_{max}=6$ is needed to ensure good convergence in the range 
considered.

\begin{figure}[tbp]
\begin{center}
\epsfig{file=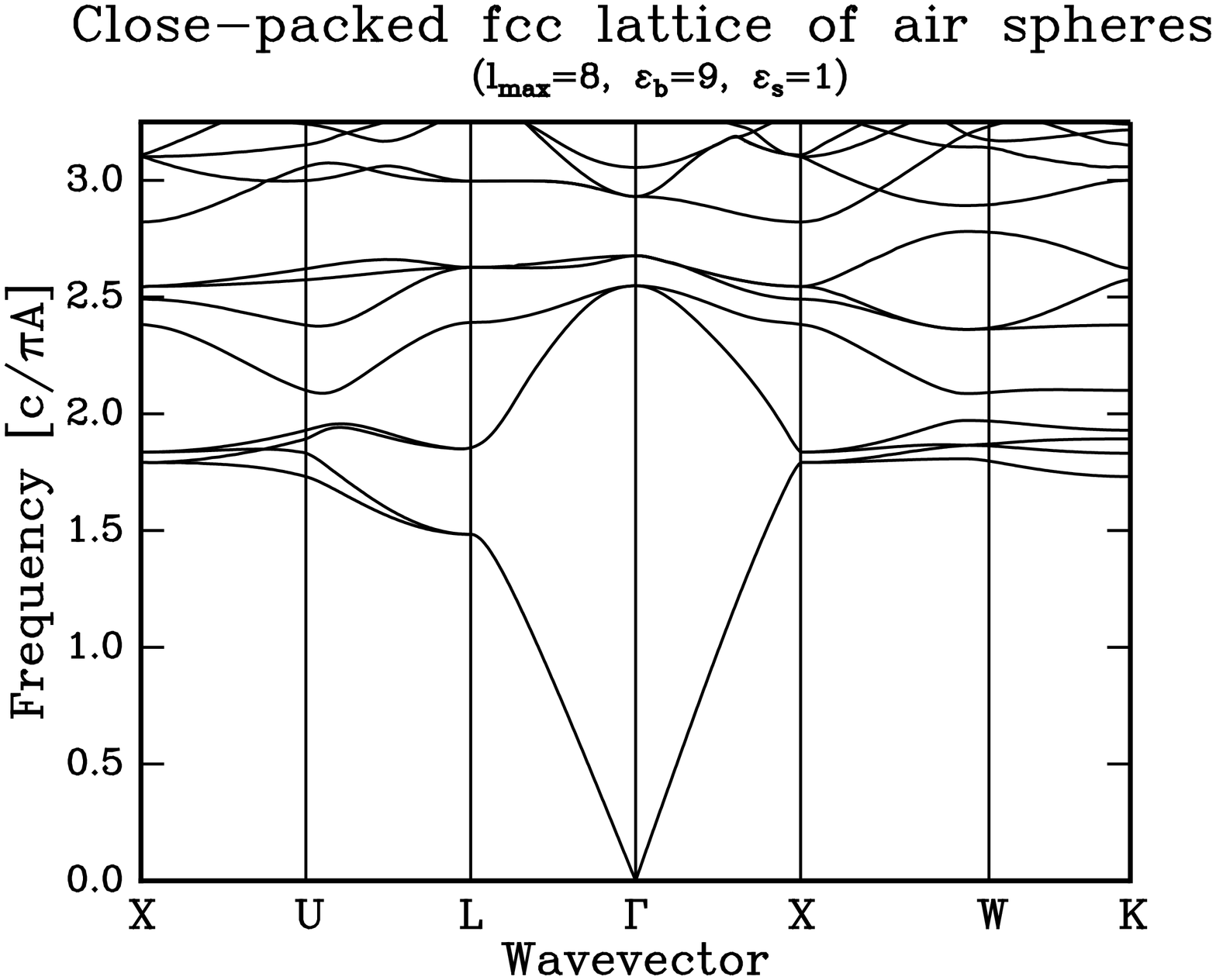,width=10cm,clip=0,angle=90}
\end{center}
\caption{Photonic band structure for a close-packed fcc lattice of
air spheres in a background 
dielectric medium with $\varepsilon_b=9$ ($n_b=3$).
Frequency is plotted in dimensionless units, $A$ is the lattice
constant, $c$ is the speed of light in the vacuum.
Only a single gap with the central gap frequency $\nu=2.796$
and the width  $\triangle\nu=0.044$ opens in the spectrum.}
\label{fg8e9}
\end{figure}

\begin{figure}[tbp]
\begin{center}
\epsfig{file=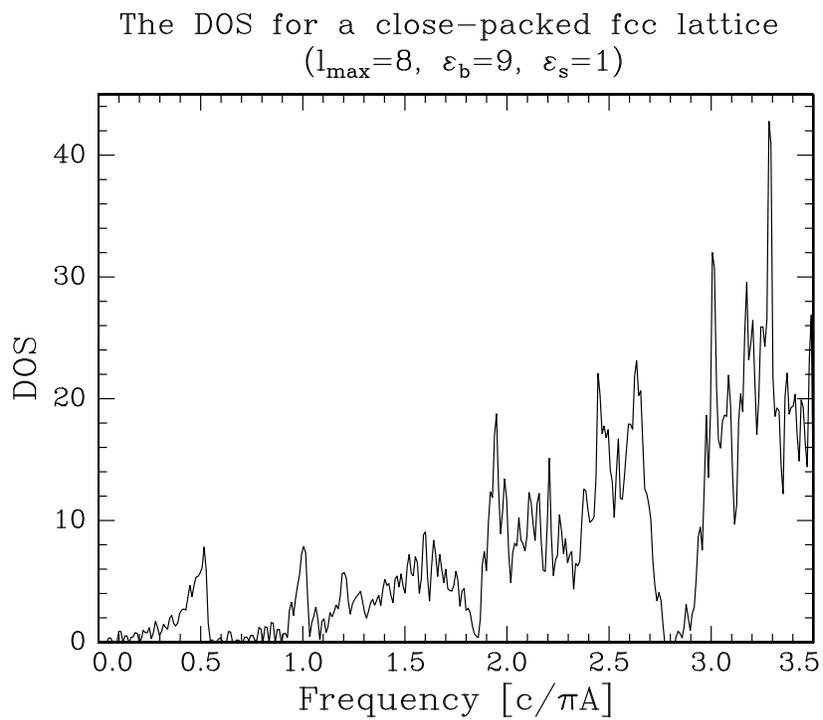,width=10cm,clip=0,angle=90}
\end{center}
\caption{The DOS per primitive cell for a close-packed fcc lattice of
air spheres in a background 
dielectric medium with $\varepsilon_b=9$ ($n_b=3$).
Note the gap in the spectrum centered at $\nu=2.796$.}
\label{fgds8e9}
\end{figure}

In figures \ref{fg8e9} and \ref{fgds8e9}
we present our results for a three-dimensional close-packed 
fcc lattice of air spheres in a dielectric medium
with the dielectric constant $\varepsilon_b=9$.
We choose this configuration for two reasons.
First, it is sufficiently representative to show
the presence of a full gap in the spectrum, and secondly,
the value of the background dielectric constant $\varepsilon_b=9$
is close to that of rutile (TiO${}_2$) at optical frequencies which
is used in experiments. 
figure \ref{fg8e9} shows the band structure.
Frequency $\nu$ is plotted in scale-invariant units $c/\pi A$, where
$A$ is the lattice constant\footnote{Note that $A$ is the side
of the conventional unit cell of the cubic lattice, which has four
times the volume of a primitive fcc unit cell,
and not the lattice spacing \cite{AsM}.}
and $c$ is the speed of light in the vacuum. 

Only a single gap with a middle of gap frequency $\nu=2.796$
and the width  $\triangle\nu=0.044$ opens in the spectrum
in the range considered. The error is determined from the convergence
properties of the KKR method.
In the close-packed case, the lower gap boundary takes on its maximal 
value at the $W$ point of the Brillouin zone while the upper gap boundary
takes on  its minimal value at the $X$ point, in agreement with the
plane-wave calculations (see \cite{Kos} for the classification of special
points of three-dimensional lattices).
In general, the photonic bands show much more  branching 
than the electronic bands and the actual classification
of different bands can be quite involved.
Group-theoretical classification of eigenmodes in three-dimensional
photonic lattices is discussed in \cite{OT}.

The presence of the gap in the $\varepsilon_b=9$, $\varepsilon_s=1$ 
case is also transparent  from the calculation of the  DOS
per primitive unit cell. The latter was calculated using the 
Monkhorst-Park integration scheme \cite{MP}. Integration over the 
Brillouin zone started from a mesh of $12\times 12\times 12$ 
uniformly spaced points, which was  subsequently reduced 
to  $182$ points with calculated weigths
using the  symmetries of the lattice.
The resulting DOS per primitive cell is plotted 
in figure \ref{fgds8e9}.

\begin{figure}[tbp]
\begin{center}
\epsfig{file=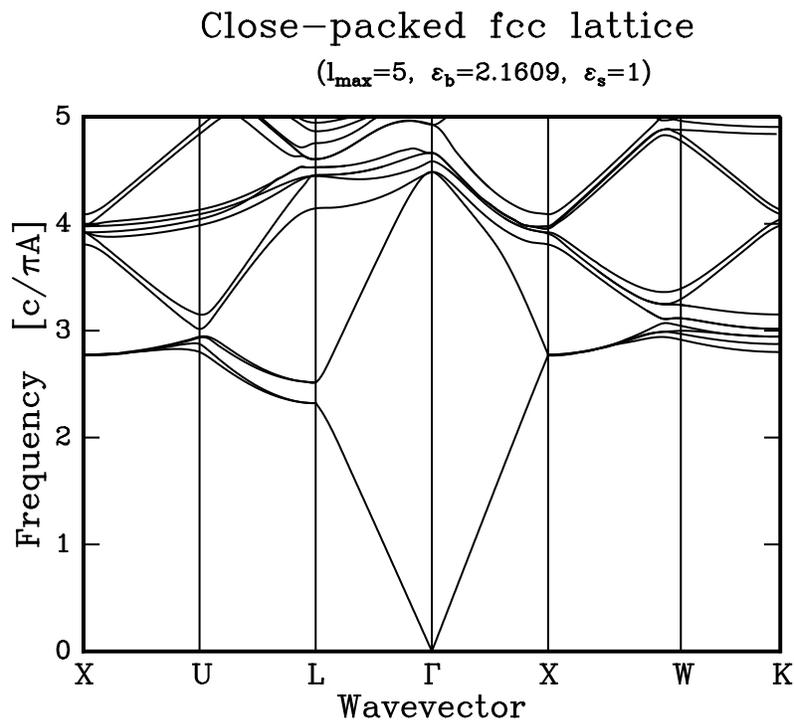,width=10cm,clip=0,angle=90}
\end{center}
\caption{Photonic band structure for a close-packed fcc lattice of
air spheres in a background dielectric medium with $\varepsilon_b=2.1609$ 
($n_b=1.47$) - experimental setup reported in [11].}
\label{fgvlas}
\end{figure}

Figure \ref{fgvlas} shows the band structure for a  close-packed 
fcc lattice of air spheres in a dielectric medium
with the dielectric constant $\varepsilon_b=2.1609$ ($n_b=1.47$). 
The latter case corresponds to the experimental setup
of \cite{As} and is also close to that of \cite{WV}.
Our calculation shows no gap in the spectrum.
Only stop gaps are present. In agreement with the
experimental observation, the most pronounced
stop gap is seen between first  bands at the $L$ point of the
Brillouin zone. For comparison with experiment, in Tab. I we give the 
width $\triangle\nu_L$ of the stop gap at the L point 
and the effective refractive index $n_{e\!f\!f}$
for a close-packed fcc lattice of air spheres in  background
media with $n_b=1.33,\, 1.37,\, 1.47,$ and  $1.6$
used in recent experiments \cite{As,WV}, together with the case $n_b=3$
for which the band structure was calculated (see figure\ \ref{fg8e9}). 
%
%
\begin{center}
TABLE I. The width $\triangle\nu_L$ of the stop gap at the L point, 
effective refractive indices $n_{e\!f\!f}$ and  $n_{e\!f\!f}^{MG}$,
and strength parameters $\varepsilon_r$ and $\Psi$  for
a close-packed fcc lattice of air spheres in different background
media. 
\vspace*{0.5cm} \\
\begin{tabular} {|c|c|c|c|c|c|} \hline\hline
 & & & & &\\
 &  $n_b=1.33$ &  $n_b=1.37$  & $n_b=1.47$ &  $n_b=1.6$ & $n_b=3$ \\
 & & & & &\\ \hline
 & & & & &\\
 $\triangle\nu_L$ &  $0.145$   &  $0.159$   &  $0.195$   & $0.236$ & 
 $0.368$
 \\
 & & & & &\\ \hline
 & & & & &\\ 
 $n_{e\!f\!f}$  &  1.084 &  1.094 & $1.120$  & $1.153$ &  $1.567$
\\
 & & & & &\\ \hline
 & & & & &\\ 
 $n_{e\!f\!f}^{MG}$  &  $1.085$   & $1.096$  & $1.122$  & $1.158$
  & $1.607$
\\
 & & & & &\\ \hline
 & & & & &\\ 
 $\varepsilon_r$  &  $0.281$   & $0.313$  & $0.391$  & $0.487$
  & $1.140$
\\
 & & & & &\\ \hline
 & & & & &\\ 
 $\Psi$  &  $-0.376$   & $-0.410$  & $-0.485$  & $-0.566$
  & $-0.935$
 \\
 & & & &  &\\
 \hline\hline
\end{tabular}
\end{center}
\vspace*{0.5cm} 
%
%
\noindent
The effective refractive index $n_{e\!f\!f}$
is determined as the inverse of the slope of the band structure around 
the $\Gamma$ point,
\begin{equation}
n_{e\!f\!f}^{-1}=\lim_{k\rightarrow 0}\, \frac{1}{c}\frac{d\omega}{dk},
\end{equation}
where $\omega$ is the angular frequency.
In the third row of Tab. I, we show the refractive index  $n_{e\!f\!f}^{MG}$ 
calculated by the Maxwell-Garnett formula \cite{MG},
\begin{equation}
n_{e\!f\!f}= \left[\varepsilon_b \,\left(
\frac{2\varepsilon_b+\varepsilon_s + 2f(\varepsilon_s-\varepsilon_b)}
{2\varepsilon_b+\varepsilon_s - f(\varepsilon_s-\varepsilon_b)}
\right)\right]^{1/2},
\label{maxg}
\end{equation}
where $f$ is the filling fraction ($f=0.7405$ for a close packed fcc 
lattice).  
In accordance with the plane-wave results \cite{DCH} (see figure 2 there),  
$n_{e\!f\!f}^{MG}$ gives the upper bound on $n_{e\!f\!f}$.
For $\varepsilon_b<\varepsilon_s$ the situation is reversed and 
$n_{e\!f\!f}^{MG}$ is expected to give the lower bound on $n_{e\!f\!f}$
\cite{DCH}. For completeness, we also show  parameters
\begin{equation}
\varepsilon_r =\left[\frac{f\varepsilon_s^2 +(1-f)\varepsilon_b^2}
{[f\varepsilon_s +(1-f)\varepsilon_b]^2} -1
\right]^{1/2},
\end{equation}
introduced in \cite{SHI},  and
\begin{equation}
\Psi = 3f\,\frac{\varepsilon_s-\varepsilon_b}{\varepsilon_s+2\varepsilon_b},
\end{equation}
introduced in \cite{WV}, which should characterize the scattering
strength of a dielectric lattice.

From the experimental point of view, it is interesting
to know what is the threshold dielectric contrast  
$\varepsilon_{m\!a\!x}/\varepsilon_{m\!i\!n}$, where  $\varepsilon_{m\!a\!x}$
($\varepsilon_{m\!i\!n}$) is bigger (smaller) of the $\varepsilon_s$ 
and $\varepsilon_b$, for which a full gap opens in the spectrum.
Obviously, this threshold value changes with the radius of spheres and 
also depends on whether the dielectric constant of spheres is larger 
or smaller than that of the background medium.
The precise value of the threshold dielectric contrast 
has been out of reach of the plane-wave calculations
\cite{SHI}. Using the photonic KKR method, we scanned
different configurations between the $X$ and $W$ points of the
Brillouin zone. 
For close-packed air spheres, the lower and upper
bounds of the full gap are set at the $W$ and $X$ points, respectively.
For smaller filling fractions, already at $f=0.70$, and close to the
threshold dielectric contrast,
the gap width is completely determined by the band structure at 
the $W$ point.  We  determined  the lowest threshold dielectric contrast 
 $\varepsilon_b/\varepsilon_s$  for an fcc lattice of dielectric spheres 
to be $8.13$ 
($\varepsilon_r=1.096$ and $\Psi=-0.918$ in this case). 
This can for example be realized for the case of close-packed air spheres 
in a background dielectric medium with the dielectric constant 
$\varepsilon_b=8.13$.
In all other cases, i.e., if the radius of spheres is lowered,
the threshold dielectric contrast is higher.
The threshold dielectric constrast  obtained by the photonic
KKR method implies the threshold refractive index constrast 
$2.8506$ which is significantly higher
than the early theoretical estimate $1.21$  by 
Yablonovitch \cite{Y} and $1.46$  by John \cite{Jo}.
On the other hand, the threshold dielectric constrast is slightly lower 
than $2.9$ obtained by the plane wave method \cite{BSS}.

\begin{figure}[tbp]
\begin{center}
\epsfig{file=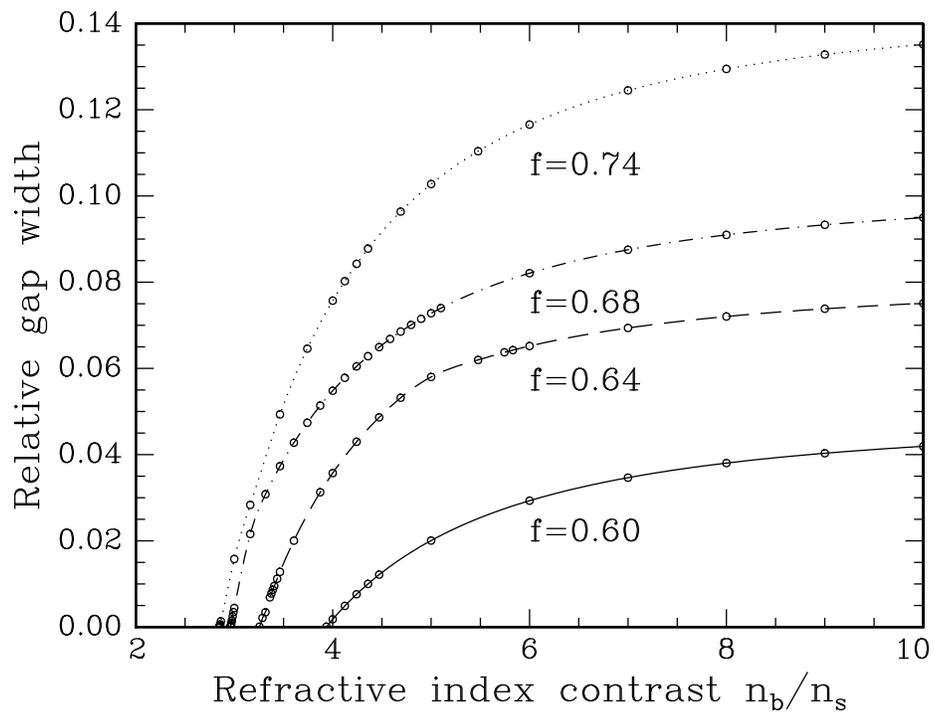,width=10cm,clip=0,angle=90}
\end{center}
\caption{Relative band-gap width, which is the band gap width divided by
the midgap frequency, as a function of the refractive index contrast
for different filling fractions $f$.}
\label{fgrww}
\end{figure}

The plot of the relative gap width, which is the gap width divided by
the midgap frequency, as a function of the the refractive index 
contrast for different filling fractions is presented in figure \ref{fgrww}.
The maximal value of the relative gap width approaches $14\%$ in the
close-packed case and  decreases monotonically 
as the filling fraction decreases. The relative gap width as a function
of the refractive index contrast shows a rapid saturation.
For example, in the close-packed case the relative gap width at $5.48$ 
is already $80\%$ of its maximal value.

\section{Face-centered cubic lattice of coated spheres}
After showing in the previous section that the photonic analogue
of the KKR method is a viable alternative to the plane-wave method
to calculate a photonic band structure, we shall proceed by
investigating the case of coated spheres.
A coated sphere is a sphere with the dielectric constant $\varepsilon_1$
and the radius  $r_1$ embedded in a larger sphere with the dielectric 
constant $\varepsilon_2$ and the radius  $r_2>r_1$, which in turn can be 
embedded in a larger sphere, etc. Let $\varepsilon_b$ be, as above, 
the background dielectric constant. 
Our interest in this system comes from the fact that there is
an intensive experimental activity to produce lattices of
coated spheres by using them as basic particles in collodial crystals.
As in the previous section, due to 
its experimental relevance, we shall investigate only a simple 
fcc lattice (one scatterer per unit cell). It turns out that a coating 
alone does not help to  create full gap extending over all the 
Brillouin zone as long as an fcc lattice
remains a simple lattice. Nevertheless, following a
suggestion by A. van Blaaderen, we shall show that a 
suitable coating of  spheres forming a dielectric lattice 
can significantly enhance (by as much as $50\%$) some of the stop gaps.
We looked mainly at the so called lowest L-gap width 
(see \cite{Kos} for the classification of special points of 
three-dimensional lattices), which corresponds to the (111) crystal direction
(see \cite{SY1} for a related theoretical discussion of the L-gap). 
The reason is that experimental techniques make it possible to allow one
to grow collodial crystals such that the L direction corresponds
to normal incidence on the crystal surface.
Another reason is that recently \cite{Mo1}, in the case of air spheres,
a simple formula has been found for the L-gap width $\triangle_L$ 
of an fcc lattice. For the case of dense spheres such an
understanding is missing. There is a hope that fcc lattices
of coated spheres could provide some insight into the L-gap behaviour.

At first it seems that a coating will not have any significant effect
on gap widths. One can arrive at this conclusion by looking
at the effective dielectric constant $\varepsilon_{e\!f\!f}^c$ for a
lattice of coated spheres. As we have seen in the previous section,
the effective dielectric constant is a measure of the scattering 
strength of a dielectric lattice almost up to the first stop gap
(Bragg reflection peak).  Let us begin with the case of a coated sphere with
a single coating, in which case we shall denote the dielectric
constant and radius of the outer shell by $\varepsilon$ and $r_s$,
respectively. Defining $x=r_1^3/r_s^3$,
\begin{equation}
\alpha_1=\frac{\varepsilon_1-\varepsilon}{\varepsilon_1+2\varepsilon},
\hspace*{1.4cm}
\alpha_0=\frac{\varepsilon-\varepsilon_b}{\varepsilon+2\varepsilon_b},
\end{equation}
the polarization factor $\alpha_c$ of a coated sphere with
a single coating is
\begin{equation}
\alpha_c =
\frac{\alpha_0+ x\alpha_1(\varepsilon_b+2\varepsilon)/
(\varepsilon+2\varepsilon_b)}
{1+2x\alpha_1\alpha_0}\cdot
\end{equation}
Here  $0\leq x\leq 1$ and $0\leq \alpha_j<1$.
The  Maxwell-Garnett formula \cite{MG} then gives
\begin{equation}
\varepsilon_{e\!f\!f}^c \approx \varepsilon_b \,
(1+ 2\, f\alpha_c)/(1- f\alpha_c),
\label{maxgsp}
\end{equation}
One can verify that for $x=1$, i.e. if the radius of the interior
sphere coincides with that of the entire sphere,
\begin{equation}
\alpha_c=\frac{\varepsilon_1-\varepsilon_b}{\varepsilon_1+2\varepsilon_b},
\end{equation}
i.e. one reproduces the  polarization factor of a homogeneous sphere
with the dielectric constant $\varepsilon_1$ in the host with 
the dielectric constant $\varepsilon_b$.
The same result one also recovers in the special case 
$\varepsilon_1=\varepsilon$.
In the limit $x\rightarrow 0$, i.e. if the radius of the interior
sphere shrinks to zero, $\alpha_c\rightarrow \alpha_0$,
the polarization factor of a homogeneous sphere with the dielectric
constant $\varepsilon$ in the host with the dielectric
constant $\varepsilon_b$.
It is important to realize that as a function of the parameter $x$,
$\alpha_c(x)$ is a monotonic function 
continuously interpolating between $\alpha_0$ and $\alpha_1$.
The latter follows from the fact that the first derivative
of $d\alpha_c(x)/dx$ has a constant sign determined by the sign of
$\varepsilon-\varepsilon_b$. Since
\begin{equation}
\frac{d\varepsilon_{e\!f\!f}^c}{d\alpha_c}= 
\frac{3f\varepsilon_b}{(1-f\alpha_c )^2} > 0,
\end{equation}
also the effective dielectric constant $\varepsilon_{e\!f\!f}^c$
is a monotonic function of $x$ which smoothly interpolates
between the two limiting cases of  homogeneous spheres
with the dielectric constants $\varepsilon_{1}$
and $\varepsilon$, respectively, in the background with 
the dielectric constant $\varepsilon_{b}$. 
Therefore, for any $x$ the effective dielectric constant 
$\varepsilon_{e\!f\!f}^c$ for a lattice of coated spheres 
is always smaller than the largest of the effective dielectric constants
obtained for the two limiting cases of a lattice of homogeneous spheres.
Using a transfer matrix method to calculate the polarization factor 
$\alpha_c$ for a coated sphere with an arbitrary
number of coatings, one can show that 
this restriction on the values of the effective dielectric constant 
$\varepsilon_{e\!f\!f}^c$ for a lattice of coated spheres  
holds also in the general case. Indeed, let us consider a coated sphere
made out of $N$ spherical shells with the dielectric
constant $\varepsilon_j$, $j=1,\ldots, N$. Then
\begin{equation}
\varepsilon_{e\!f\!f}^c \leq  \max_j \{\varepsilon_{e\!f\!f}^j \}
\end{equation}
where $\varepsilon_{e\!f\!f}^j$ is the effective dielectric constants
of a lattice of homogeneous spheres with the dielectric
constant $\varepsilon_j$ in the background $\varepsilon_b$.
Therefore, one would naively expect that a coating has mere effect
of interpolating between the limiting cases of homogeneous spheres.

\begin{figure}[tbp]
\begin{center}
\epsfig{file=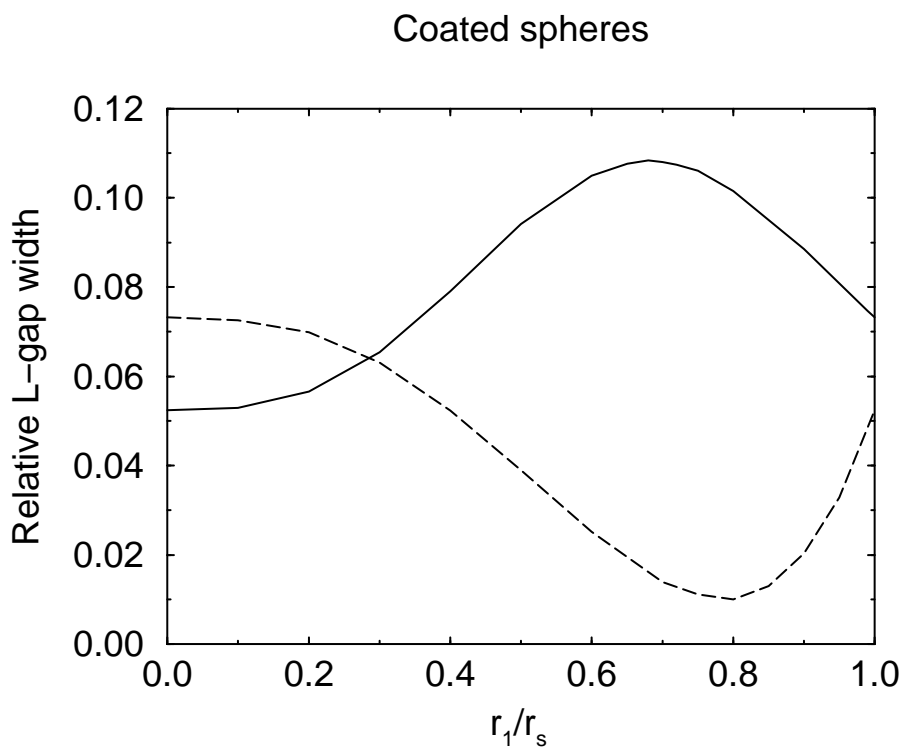,width=10cm,clip=0,angle=-90}
\end{center}
\caption{The relative L-gap width for a simple closed packed
fcc lattice of coated spheres as a function of the ratio $r_1/r_s$
of the interior and whole sphere radii. The solid line corresponds
to the case when the refractive index of the core sphere is  $n_1=2$ 
(ZnS)
and that of the shell is $n=1.45$ (silica). The dashed line shows
the reversed case, i.e., $n_1=1.45$ and $n=2$. Note that in the first
case the relative L-gap width can be increased by as much as 50\% 
if $r_1/r_s \approx 0.68$. The cases $r_1/r_s=0$ and $1$ correspond
to the limiting case of homogeneous spheres.}
\label{fgcpcslg}
\end{figure}

Nevertheless, considerations based on the value of $\varepsilon_{e\!f\!f}^c$
can be deceptive. 
In figure\ \ref{fgcpcslg} we show a plot of the relative
L-gap width $\triangle_L$ for a close-packed fcc lattice of coated spheres
as a function of $r_1/r_s$, the ratio of the interior and whole 
sphere radii.  The solid line corresponds
to the case when the refractive index of the core sphere is  $n_1=2$ 
(ZnS) and that of the shell is $n=1.45$ (silica). 
The dashed line shows the reversed case, i.e., $n_1=1.45 < n=2$. 
Obviously, the cases $r_1/r_s=0$ and $1$ correspond
to the limit of homogeneous spheres. Let us denote by 
$\triangle_+$  the larger relative L-gap width of the two 
limiting homogeneous spheres cases.
Then in the first case ($n_1 > n$) around  $r_1/r_s \approx 0.68$ 
the relative L-gap width can be increased by  as much as 50\%
with respect to $\triangle_+$.
In the second case ($n_1 < n$), the relative
L-gap width remains smaller than 
$\triangle_+$ for all values of $r_1/r_s$.

In the following, we tested the Maxwell-Garnett formula 
(\ref{maxgsp}) against  the exact value of the effective dielectric 
constant $\varepsilon_{e\!f\!f}^c$ (refractive index 
$n_{e\!f\!f}=\sqrt{\varepsilon_{e\!f\!f}^c}$)
obtained directly from a band structure.
In Table~II we collected the  effective refractive indices 
$n_{e\!f\!f}$ and  $n_{e\!f\!f}^{MG}$ (the Maxwell-Garnett value)
for those values of $r_1/r_s$ which are nearby the maximum and
minimum of the relative L-gap width $\triangle_L$
(cf. figure\ \ref{fgcpcslg}).
Here one expects the largest deviations of $n_{e\!f\!f}^{MG}$
from $n_{e\!f\!f}$ because $n_{e\!f\!f}^{MG}$
is a monotonic function of $r_1/r_s$ and hence, it does not have
local minima and maxima.
%
%
\begin{center}
TABLE II. The  effective refractive indices $n_{e\!f\!f}$ and  
$n_{e\!f\!f}^{MG}$
for a simple close-packed fcc lattice of coated  spheres in air. 
First three columns are for $n_1=2$ and $n=1.45$, the last three
colums are for the reversed case  $n_1=1.45$ and $n=2$.
\vspace*{0.5cm} \\
\begin{tabular} {|c||c|c|c||c|c|c|} \hline\hline
 & & & & & &\\
 $r_1/r_s$ &  $0.65$ &  $0.68$  & $0.72$ &  $0.75$ & $0.8$ & $0.85$ \\
 & & & & & &\\ \hline
 & & & & & &\\
 $n_{e\!f\!f}$  &  1.419 &  1.434 & $1.444$  & $1.540$ &  $1.507$ &  $1.469$
\\
 & & & & & &\\ \hline
 & & & & & &\\ 
 $n_{e\!f\!f}^{MG}$  &  $1.414$   & $1.428$  & $1.438$  & $1.525$
  & $1.494$ & $1.460$
 \\
 & & & & & &\\
 \hline\hline
\end{tabular}
\end{center}
\vspace*{0.5cm} 
%
It is interesting to note that $n_{e\!f\!f}^{MG}$ continues to
approximate the exact value $n_{e\!f\!f}$ very well.
Similarly as for the case of homogeneous spheres in air \cite{DCH},
in the case of coated spheres in air the 
Maxwell-Garnett formula (\ref{maxgsp}) slightly {\em underestimates}
the exact value of the  effective refractive index $n_{e\!f\!f}$
obtained from a band structure
\begin{equation}
n_{e\!f\!f}^{MG} < n_{e\!f\!f}.
\end{equation}

\section{Conclusions}
We showed that the photonic KKR method is a viable alternative 
to calculate the photonic band structure by  reproducing the main 
features of  the  photonic band structure  obtained by the plane-wave method
for a simple three-dimensional fcc lattice of homogeneous spheres 
\cite{SHI,SY,BSS}. We confirmed that for a sufficiently high dielectric 
contrast a full gap opens  between the eights and ninth bands,
a direct gap opens between the fifth and sixth bands,
and no gap opens in the spectrum if $\varepsilon_b <\varepsilon_s$.
To obtain a good convergence in the frequency range considered,
it was sufficient to retain multipole fields  with the angular momentum 
up to $l_{max}=6$. In general, the higher frequency the higher 
value of  $l_{max}$ is needed. In order to reproduce the first
band and the linear part of the spectrum, $l_{max}=1$  is enough. 
The size of a secular equation is reduced by almost a factor $10$
compared  the plane-wave method \cite{ZS}-\cite{BSS} which 
customarily requires well above thousand plane waves. 
Precision of the elements
of the secular equation is determined by the standard Ewald 
summation \cite{KKR} which yields structure constants \cite{Mo} up to 6 digits.

For close-packed air spheres, the lower and upper
bounds of the full gap are set at the $W$ and $X$ points, respectively.
For filling fractions less than $0.7$ and close to the
threshold dielectric contrast for which a full gap opens in the spectrum,
the gap width is completely determined  at the W point.
The lowest dielectric contrast $\delta=\varepsilon_b/\varepsilon_s$ 
for which a full gap  opens in the spectrum is found to be $8.13$ which occurs
for a close-packed fcc lattice. This value is slightly lower than $8.4$
obtained by the plane-wave method \cite{BSS}.
The maximal value of the relative gap width approaches $14\%$ in the
close-packed case and  decreases monotonically 
as the filling fraction decreases. 
For readers interested in the the role of Mie's resonances
in the formation of a band structure we refer to \cite{MT1}.

\begin{figure}[tbp]
\begin{center}
\epsfig{file=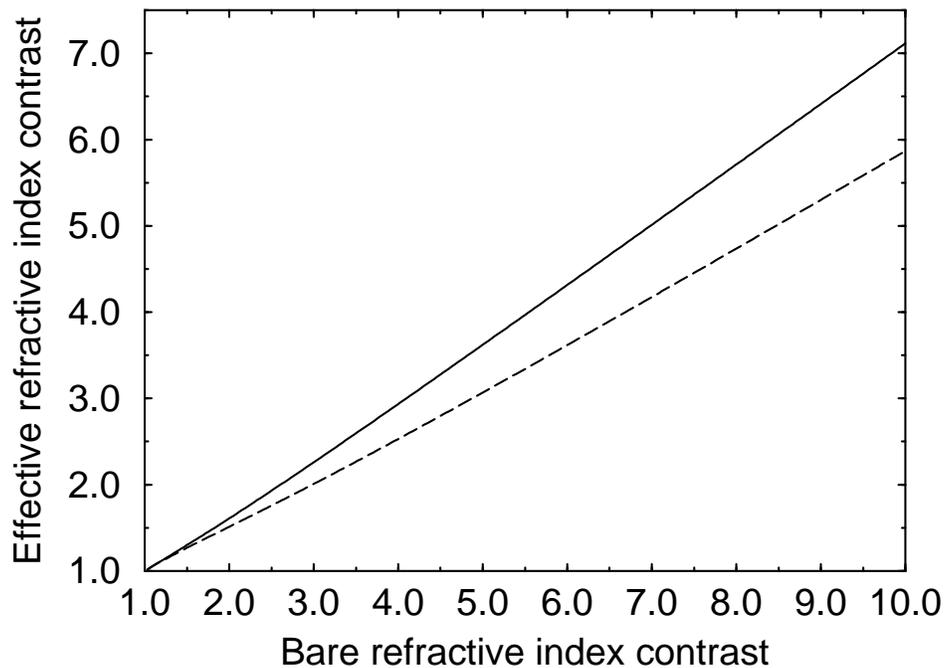,width=10cm,clip=0,angle=-90}
\end{center}
\caption{In a medium with finite density of scattering spheres,
such as for spheres on a lattice, the bare refractive index contrast
gets renormalized. This may give a partial explanation
of the fact why the band structure of an fcc lattice 
of the dense and air spheres shows such a different behaviour.
This figure shows
a typical behaviour of the effective refractive index contrast
as a function of the  bare refractive index contrast
for air spheres in a dielectric (solid line) and dense 
dielectric spheres in air (dashed line). Note that the latter 
is always smaller than the former. This plot was made for
filling fraction $f=0.4$.
The Maxwell-Garnett overestimates $\varepsilon_{e\!f\!f}$
for the case of air spheres and underestimates $\varepsilon_{e\!f\!f}$
for the case of dielectric spheres in air.
As the result, the exact curves are rotated slightly to the right
with respect to the case when the Maxwell-Garnett value is taken
for $\varepsilon_{e\!f\!f}$.
}
\label{contrff}
\end{figure}

An open question remains what causes such a different behaviour
of a lattice of dense spheres compared to that of air spheres. 
This can be partially attributed to the  fact that, 
for a lattice of spheres, $\varepsilon_b$ no 
longer describes the dielectric constant of the surrounding medium, 
which is instead described by the effective dielectric constant
$\varepsilon_{e\!f\!f}$. Therefore, the bare dielectric contrast 
$\delta$  is renormalized to 
$\delta_{e\!f\!f}=\max (\varepsilon_s/\varepsilon_{e\!f\!f},
\varepsilon_{e\!f\!f}/\varepsilon_s)$, where 
$1 < \varepsilon_{e\!f\!f} < \delta$ for  
$\varepsilon_s\neq \varepsilon_b$.
Given the bare dielectric contrast $\delta$, one finds that
the  renormalized dielectric contrast 
$\delta^d_{e\!f\!f}=\varepsilon_s/\varepsilon_{e\!f\!f}$ in the case 
of dense spheres is 
always smaller than the  renormalized dielectric contrast 
$\delta^a_{e\!f\!f}=\varepsilon_{e\!f\!f}$
in the case of air spheres (see figure \ref{contrff}). 
The latter is easy to verify in the limit of very high 
bare dielectric contrast $\delta\rightarrow \infty$, 
where Maxwell-Garnett's formula (\ref{maxgsp}) implies
$$
\delta_{e\!f\!f}^d \sim \delta \,(1-f)/(1+2f)
< \delta_{e\!f\!f}^a \sim \delta\,(1-f)/(1+f/2).
$$
Nevertheless, a full understanding of the differences between the
lattices of air and dense spheres still remains a theoretical
challenge.

Such as in the case of homogeneous spheres in air, also in the case of a 
simple fcc lattice of coated spheres in air 
no full gap opens in the spectrum. However, we showed
that already a single coating can enhance the relative L-gap width 
$\triangle_L$ by as much as 50\% with respect to $\triangle_+$,
the larger relative L-gap width of the two limiting cases of homogeneous 
spheres. This suggests  the use of coated spheres to reduce the threshold 
dielectric contrast for which a full gap opens in 
the spectrum in the case of the so-called complex fcc lattice
(having more than one scatterer in the primitive cell),
which will be dealt with elsewhere.
Note that in contrast to a simple fcc lattice, in the latter case a full 
gap does open in the spectrum \cite{BSS}.
In general, $\triangle_L$ is not a monotonic function of the effective 
refractive index $n_{e\!f\!f}$ and, for a single coating, shows one 
local maximum (minimum) as $r_1/r_s$ is varied from
zero to one (cf. figure\ \ref{fgcpcslg}).

It is interesting to note that the Maxwell-Garnett formula (\ref{maxgsp})
provides a very good approximation to the exact value of the 
effective refractive index $n_{e\!f\!f}$ also for coated spheres.
Similarly to the case of homogeneous spheres in air \cite{DCH},
in the case of coated spheres in air the 
Maxwell-Garnett formula \cite{MG} slightly {\em underestimates}
the exact value of the  effective refractive index $n_{e\!f\!f}$
obtained directly from a band structure.
The Maxwell-Garnett formula holds well beyond the limit
$\omega r_s \ll 1$, $r_s$ being the sphere radius,
for which it was originaly derived.
It describes $n_{e\!f\!f}$ well almost up to the lowest L gap
and can also be used to determine the L-midgap 
angular frequency $\omega_c$ \cite{Mo1}. 
In the case of coated spheres considered here 
(cf. figure \ref{fgcpcslg}), 
the ratio $ n_{e\!f\!f} \omega_c /k_L \in (0.973,0.995)$,
where $k_L$ is the length of the Bloch vector
corresponding to the L point.

\section{Acknowledgments}
A.M. wishes to thank A. Polman and A. Tip  for careful reading of the 
manuscript and discussions, A. van Blaaderen for his suggestion 
on coated spheres, and M.J.A. de Dood for help with plots. This work is
part of the research program by  the Stichting voor Fundamenteel Onderzoek 
der Materie  (Foundation for Fundamental Research on
Matter) which  was made possible by financial support 
from the Nederlandse
Organisatie voor Wetenschappelijk Onderzoek (Netherlands Organization for
Scientific Research).  SARA computer facilities are also gratefully
acknowledged.
C.S. wants to acknowledge Idris computing center.

\newpage

\end{document}